\documentclass[sigconf]{acmart}

\AtBeginDocument{%
  \providecommand\BibTeX{{%
    \normalfont B\kern-0.5em{\scshape i\kern-0.25em b}\kern-0.8em\TeX}}}

\setcopyright{acmlicensed}
\copyrightyear{2024}
\acmYear{2024}
\acmDOI{XXXXXXX.XXXXXXX}

\acmConference[KDD '24]{the 30th ACM SIGKDD Conference on Knowledge Discovery and Data Mining, Aug 25--29,
  2024, Barcelona, Spain}{Aug 25--29,
  2024}{Barcelona, Spain}
\acmISBN{978-1-4503-XXXX-X/18/06}

\usepackage{listings}
\usepackage{longtable}
\usepackage[export]{adjustbox}

\begin{document}
	
\title{Chaining text-to-image and large language model: A novel approach for generating personalized e-commerce banners}

\author{Shanu Vashishtha}
\email{shanu.vashishtha@walmart.com}
\affiliation{%
  \institution{Walmart, Inc.}
  \city{Sunnyvale, CA}
  \country{USA}}

\author{Abhinav Prakash}
\email{abhinav.prakash@walmart.com}
\affiliation{%
  \institution{Walmart, Inc.}
  \city{Sunnyvale, CA}
  \country{USA}}

\author{Lalitesh Morishetti}
\email{lalitesh.morishetti@walmart.com}
\affiliation{%
  \institution{Walmart, Inc.}
  \city{Sunnyvale, CA}
  \country{USA}}

\author{Kaushiki Nag}
\email{kaushiki.nag@walmart.com}
\affiliation{%
  \institution{Walmart, Inc.}
  \city{Sunnyvale, CA}
  \country{USA}}

\author{Yokila Arora}
\email{yokila.arora@walmart.com}
\affiliation{%
  \institution{Walmart, Inc.}
  \city{Sunnyvale, CA}
  \country{USA}}

\author{Sushant Kumar}
\email{suhsant.kumar@walmart.com}
\affiliation{%
  \institution{Walmart, Inc.}
  \city{Sunnyvale, CA}
  \country{USA}}

\author{Kannan Achan}
\email{kannan.achan@walmart.com}
\affiliation{%
  \institution{Walmart, Inc.}
  \city{Sunnyvale, CA}
  \country{USA}}

\renewcommand{\shortauthors}{Vashishtha et al.}

\begin{abstract}
Text-to-image models such as stable diffusion have opened a plethora of opportunities for generating art. Recent literature has surveyed the use of text-to-image models for enhancing the work of many creative artists. Many e-commerce platforms employ a manual process to generate the banners, which is time-consuming and has limitations of scalability. In this work, we demonstrate the use of text-to-image models for generating personalized web banners with dynamic content for online shoppers based on their interactions. The novelty in this approach lies in converting users’ interaction data to meaningful prompts without human intervention. To this end, we utilize a large language model (LLM) to systematically extract a tuple of attributes from item meta-information. The attributes are then passed to a text-to-image model via prompt engineering to generate images for the banner. Our results show that the proposed approach can create high-quality personalized banners for users.
\end{abstract}

\ccsdesc[500]{Applied computing~Online shopping}
\ccsdesc[500]{Computing methodologies~Computer vision}
\ccsdesc[500]{Computing methodologies~Natural language generation}
\keywords{Large Language Models, Stable Diffusion, Prompt Engineering, E-commerce, Personalization, Image Generation}



\maketitle

\section{Introduction}

Generative artificial intelligence (generative AI) has shown promising results in many applications ranging from chatbots, personal assistants, content creation, and digital art \cite{cao2023comprehensive}.  Oftentimes, the quality of results obtained through these models is at par or better than human-created content. This rapid innovation in generative AI technology has motivated businesses to invest in the acquisition, development, and deployment of these technologies for their customers. As per a Gartner poll \cite{Gartner2023} of more than 2500 executives, 38\% of the executives indicated that the primary focus of their generative AI investment is to retain customers and improve customer experience. For an e-commerce business, personalizing their web content pertaining to a user’s interests creates tremendous value for the business as well as improves the customer experience.  

The existing process to create banners for e-commerce platforms often relies on a manual process, where a creative team builds the banners for a given set of users who will be interested in a certain product category (user cohorts). This process is time-consuming and does not scale well as the number of cohorts grows. To this end, we present a novel approach to personalize banners on an e-commerce platform based on a user’s item interactions, where the items are the products being sold on the e-commerce platform.  Item interaction data contain useful information about a user's preference. Hence, the idea of using item information instead of user cohorts for creating banners provides an opportunity to generate more personalized and diverse banners. Ultimately, the user cohorts also take into account user-item interactions, so the item-level generation can be considered as a more granular banner personalization. A user may interact with multiple items in a given time frame. To resolve conflicts and select a single item among a set of eligible items for a user, we utilize the existing user cohort affinities and select the item matching the cohort for which the user has the highest affinity. Subsequently, this item is used for banner generation. 

In this work, we propose an automated banner generation process from a given item information by harnessing the capabilities of two modalities of generative AI models---large language model (LLM) and text-to-image model (Stable Diffusion).  The stable diffusion models have shown promising results in creating digital art given text prompts. However, generating a prompt that is relevant to the task at hand for the image generation model is a crucial step for the success of such models. Thus, to achieve scalability, automating prompt engineering for banner generation rather than curating a manual prompt for each user cohort becomes an important step. To solve this problem, we propose a chaining technique for LLMs with Stable diffusion models.  

Our proposed framework uses an LLM to extract a tuple of attributes/keywords from product names. These attributes denote aspects of the product that are helpful to create an image generation prompt, such as an action that can be performed using the item, or an environment or a setting in which the object is most likely to be used. The attributes are then passed on to the image generation model to create engaging banners for the users. 

We evaluate our proposed framework on multiple dimensions: i) the quality of the generated images using the BRISQUE metric; and ii) the evaluation of suitability of the generated image for the desired use case using human annotation. 

We summarize the contributions of this work as follows: 
\begin{itemize}
    \item Methodology: we developed a technique for chaining an LLM with a Stable diffusion model to automate the process of image generation for web banners. 
    \item Evaluation: we do extensive evaluation studies to compare and contrast the proposed method with other possible solutions. 
\end{itemize}

\section{Background and Related Work}
\subsection{Image Generation}
The solutions to the image generation problem have evolved over the last decade \cite{cao2024survey}. There are various methods for solving the image generation problem, such as variational autoencoders (VAEs) \cite{kingma2013auto,huang2018introvae} or generative adversarial networks (GANs) \cite{goodfellow2014generative,mao2021generative}. The VAE approach uses an encoder-decoder architecture where the encoder embeds the input image into a lower dimension latent vector space and the decoder reconstructs the image back from its latent vector representation. VAE models are known to generate diverse images; however, one of the drawbacks of VAE, as noted in the literature, is their inability to generate sharp images.   

The GAN approach, on the other hand, relies on two neural network models acting as generators and discriminators, respectively. The two networks compete against each other---the generator tries to generate as realistic an image as possible and the discriminator tries to detect fake images. In the process, both networks learn their weights and eventually become better at their respective tasks. Unlike VAE, GANs have overcome the problem of blurry images and generate sharp photo-realistic images, particularly for human faces. However, they are not able to generate a diverse set of images.  

A more recent framework that has gained popularity among researchers and practitioners alike is the diffusion model \cite{ho2020denoising}. In particular, a variant of the diffusion model known as the stable diffusion (SD) model \cite{rombach2022high} has shown promising results \cite{dhariwal2021diffusion}. Stable diffusion models build upon the VAE architecture by encoding the images to a latent vector space and then reconstructing the images using a decoder.  

Stable Diffusion-based image generation models have propelled an explosion in creativity. The active community on various social media websites like Reddit \cite{reddit2022} and Discord have experimented and pooled their knowledge to come up with the best prompting techniques to generate some amazing AI-generated content. One of the challenges in using a text-to-image generation model is prompt engineering. The prompt needs to capture user intent in a way that is comprehensible by the generation engine. In this work, we use stable diffusion as the image generation engine of the banners.

\subsection{Large Language Models}
Generative AI models such as PaLM \cite{chowdhery2023palm}, GPT-3 \cite{brown2020language}, BLOOM \cite{workshop2022bloom}, and LLAMA \cite{touvron2023llama} fall under the category of Large Language Models (LLM) that are commonly employed to generate text. LLMs can solve many of the tasks associated with natural language processing (NLP), such as text classification \cite{liang2022holistic}, natural language inference \cite{qin2023chatgpt}, and semantic understanding \cite{tao2023eveval}.

Over the last few years, we have seen an increase in the number of large language models with ever-increasing emergent abilities \cite{wei2022emergent}. These emergent abilities make them useful for tasks that they were not initially pre-trained for with little to no training (few-shot and zero-shot learning). 

Language models have also shown promising results for attribute extraction tasks \cite{wang2020learning, maragheh2023llm}. One of the key aspects of our work is extracting meaningful attributes for a product name to create a prompt for the text-to-image model. Hence, we use an LLM as our attribute extraction engine.

\subsection{Prompt Engineering for Image Generation}
Literature has proposed different ways to improve prompt engineering. Promptify \cite{brade2023promptify} developed an interactive prompt exploration tool that helps reduce the cognitive load of prompt engineering among its users. 

A recent literature survey \cite{gu2023systematic} provides an in-depth review of different prompting strategies for the text-to-image model. The survey classifies different prompting strategies: i) semantic prompt design; ii) diversify generation with prompt; and iii) complex control of synthesis results. Out of the three strategies, semantic prompt design states the key importance of objects (nouns), attributes (adjectives), and actions (verbs) to improve the alignment between the prompt (intent) and the generated images (outcome).

In this work, we focus on semantic prompt design for banner image generation as explained in detail in the next section.
\section{Methodology}

\begin{table}
\centering 
\caption{Number of words in product names for the online catalog}\label{wl_table} 
\begin{tabular} { c  c } \hline
\textbf{Metric} & \textbf{Value} \\ \hline 
Mean & 15.78 \\ 
Std. dev. & 6.35 \\
Min & 1 \\ 
Max & 265 \\ \hline
\end{tabular} 
\end{table} 

\begin{figure}
    \centering
    \includegraphics[width=\linewidth]{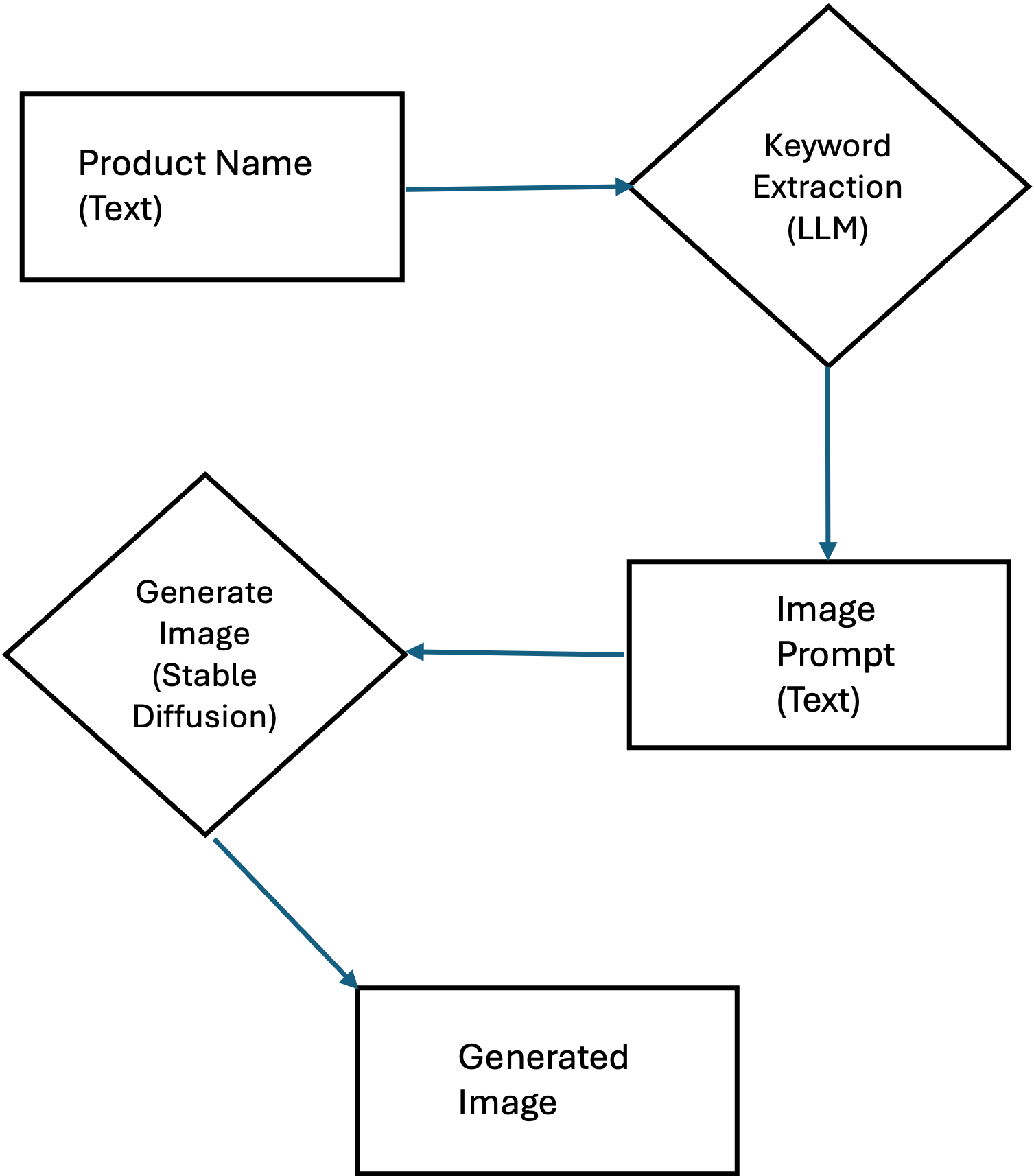}
    \caption{A flow diagram for the proposed method; rectangle denotes input/output and rhombus denotes action.}
    \label{fig:flow_diag}
\end{figure}

The problem that we are solving is to generate personalized banners and creatives for a website's welcome page or home page based on user preferences. For a small group of users, one can employ content creators and provide them with relevant backgrounds about users or groups of users with specific personas to create personalized content. However, this problem becomes challenging as we scale and aim for a granular level of personalization based on each user’s historical data. Present-day e-commerce websites also employ a more thematic update following changing seasons or holidays requiring a solution that can generate images dynamically. 

To solve this industrial application problem, we designed a novel approach for generating personalized web banners for e-commerce websites at scale by leveraging the capabilities of a large language model and an image generation model. The first step in the proposed method is to extract attributes from product names. The second step is to convert these attributes to a prompt for a text-to-image model and feed the prompt to the model to generate the banner image. Overall, our approach can be summarized using the flowchart in Figure\ref{fig:flow_diag}. At the core of the proposed method lies three sub-problems that we address:

\subsubsection{Item Representation Problem} The first sub-problem requires us to develop an effective method for interpreting and understanding the product name, for which we use a large language model (LLM). 

Our LLM-based approach involves the extraction of three key attributes from a given product name as the representation to provide in the image prompt: the actual product or the subject, some specific features that can describe the product (keywords), and a setting where the item is most likely to be used. The idea behind extracting this representation is to make the prompt for the image generation model to be clear, concise, consistent, and grounded in the domain knowledge of the LLM.

Another straightforward method to represent items is to utilize product names as inputs. However, item names can be very long with lots of details, or sometimes they might be very short and vague, which can make it challenging to extract meaningful information. In Table~\ref{wl_table}, we present summary statistics of our item catalog. Item names on average consist of 16 words with a maximum value of 256 and a minimum of 1. This distribution spread presents a challenge to only extract meaningful keywords from a given product name. However, to test the image generation capability from a product name, we did a comparison study where we also generated images by providing product names as input prompts.

\subsubsection{Image Generation Problem} The second sub-problem is tackled using a stable diffusion model capable of generating an image that highlights the product in the determined setting and showcases it with the highlighted features. This requires the development of an image generation prompt that is guided by the tuple produced by the large language model as input and generates an image as output. Given that an e-commerce catalog contains millions of items dispersed across scores of categories and thousands of sub-categories, a fixed template that works for all items is required to automate the process. The LLM creates this image-generation prompt as part of its final output.   

\subsubsection{Personalization Problem} The third sub-problem is to personalize the generated images based on the user's interests. The proposed approach can generate banners given an item's attributes/keywords. We would need to map the relevant items to users. An easy way to do so is using the item that a user has most recently interacted with. However, for any reasonable time frame, a user may interact with multiple items, and not all of those items may be relevant to them. Hence, we need a better approach to map items to users.  

In an e-commerce setting, user interests are often captured by grouping the users into cohorts. These cohorts are based on historical user-item interactions, for example, a user who frequently interacts with and transacts pet products would be assigned a ``pet-owner" cohort and a corresponding affinity score. To solve the item-to-user mapping problem for personalization, we utilize the existing user cohort affinities. Among all the candidate items available for a given user, we select the item matching the cohort for which the user has the highest affinity. This mapping ensures that we generate multiple personalized web banners (based on different items), and yet create a personalized experience that resonates with the users belonging to a particular cohort. Our future plans encompass transitioning towards a session-based user-to-product mapping framework. 

\section{Evaluation Study}

\begin{figure*}[htbp]
    \centering
    \includegraphics[width=\textwidth]{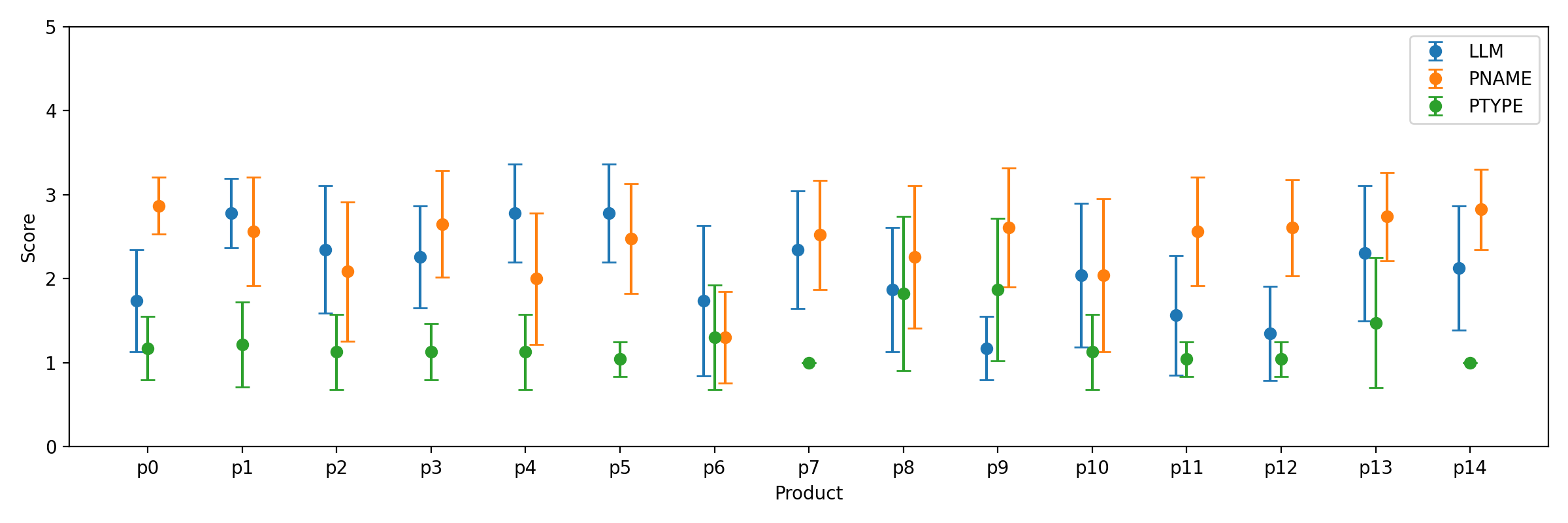}
    \caption{Average user score along with standard error for different products.}
    \label{fig:score_plot}
\end{figure*}

We evaluate our methodology using existing open-source foundation models. For prompt generation, we decided to use LLAMA 2 \cite{touvron2023llama2}. For image generation, we experimented with Stable Diffusion models (SD) – v1.5 and XL \cite{podell2023sdxl}. The evaluation is done on two important aspects: the quality of the image and its relevance to the underlying product. 

\subsection{Image quality evaluation}
Our empirical observations on image generation using SD v1.5 are consistent with those of the wider community. Images with human body parts, especially with facial features and hands are distorted and out of proportion. Images with text in them are illegible or incoherent. Since we aim to create a marketing banner, placing text in the image is one of our desired requirements. However, most prompt variations resulted in images with blurred or jumbled text. We hypothesized that the product names, which we directly provided as input prompts, being overly verbose misled the model to focus on less important keywords and lose track of the main subject being referenced. For example, \texttt{“Nisrada Calming Donut Dog Bed for Small to Large Dogs, Cat \& Puppy Bed: Anti-Anxiety, Self-Warming, Cozy Soft Plush Round Pet Bed, Ideal for Both Home \& Travel, Washable(30")”} created an image of a dog but without a Pet Bed which is the main product. We fine-tuned the text encoder for SD-v1.5 on our corpus of data which led to a slight improvement in the image generation adherence to the provided prompt.   

Since the Stable Diffusion XL model had an improved text encoder and provided images with higher resolution, we decided to switch to the Stable Diffusion XL model instead of continuing with the fine-tuning approach. To evaluate the quality of images, we used a widely used metric called BRISQUE \cite{mittal2012no}. The mean of BRISQUE scores for the generated images using the v1.5 model was 29.63. The results surpassed those garnered using the XL model, which typically yielded values within the 15-18 range. However, given the length complexity of product names, we decided to continue with the chaining LLM approach to create prompts for the stable diffusion model. 

For experimentation purposes, we sampled some product types that are popular among users. For the sampled product types, we randomly sampled item names from the catalog of items belonging to that product type. As an example, for users interested in pet products, one of the sampled product types was "pet beds". For the sampled set of item names, we experimented with three approaches of input prompts for the SD model:   

\begin{itemize}
    \item \texttt{LLM}: providing the image prompt generated from an LLM as input for image generation. In this approach, we provided the product names first to an LLM model. The LLM model is tasked with a two-step guide to create the final image generation prompt. In the first step, the LLM model extracts the product’s subject, keywords describing the product along with the setting in which the product is most used. This tuple is used to create the final prompt on the following structure - "subject" with "keywords" in "setting". The output from this LLM is provided as input to the stable diffusion model. We provide the prompt used for the LLM in Appendix \ref{llm-prompt}.  
    \item \texttt{PNAME}: providing product names directly as prompt for image generation. In this approach, we provided the sampled product names as input to the SD model for image generation.
    \item \texttt{PTYPE}: providing the product type of the given product as a prompt for image generation. In this approach, we provided the product type (a hierarchical classification in the product catalog) which captures an important level of granularity for the product. 
\end{itemize}
  
For the three different approaches mentioned, we evaluated the image quality using BRISQUE. The results are presented in Table \ref{brisque}. Given that the obtained scores were low, we maintained a significant degree of assurance regarding the quality of the generated images.

\begin{table}
\centering 
\caption{BRISQUE metric for different prompting approaches}\label{brisque} 
\begin{tabular} { c  c  c } \hline
\textbf{Input prompt} & \textbf{Mean} & \textbf{Std. dev.} \\ \hline 
\texttt{LLM} & 18.27 & 11.69 \\ 
\texttt{PNAME} & 15.45 & 11.80 \\ 
\texttt{PTYPE} & 20.77 & 17.76  \\ \hline
\end{tabular} 
\end{table} 

To quantify adherence to the input prompt, we utilized YoloV8 \cite{Jocher_Ultralytics_YOLO_2023} for obtaining the Prompt Adherence Recall (PAR) score \cite{gani2023llm}. This score is defined as the mean of object presence over all objects over all prompts to check if the object is present in the generated image such that object presence is 1 if present and 0 otherwise. However, the detector's output probabilities demonstrated a deficiency in product-specific vocabulary, which was crucial for accurately assessing the prompts. Consequently, we developed a human assessment survey specifically designed for the evaluation of the relevance of generated images.

\subsection{Human evaluation}

\begin{table*}[htbp]
	\centering
	\caption{Table with product name and generated images with different approaches }
	\label{tab:images_text}
	\begin{tabular}{p{0.195\textwidth} c c c}
		\hline
		Product Information & \texttt{LLM} & \texttt{PNAME} & \texttt{PTYPE} \\
		\hline
		Vibrant Life Luxe Cuddler Mattress Edition Dog Bed, Medium, 27"x21", Up to 40lbs &  \raisebox{-\totalheight}{\includegraphics[width=0.2\textwidth]{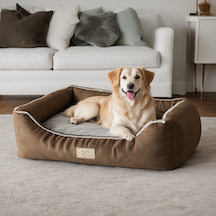}} & \raisebox{-\totalheight}{\includegraphics[width=0.2\textwidth]{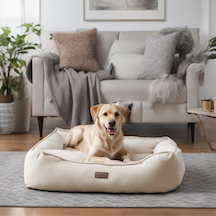}}  & \raisebox{-\totalheight}{\includegraphics[width=0.2\textwidth]{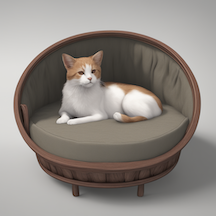}} \\
		Walker Edison 32" Scandinavian 2 Door Accent Cabinet - Coastal Oak/ Black &  \raisebox{-\totalheight}{\includegraphics[width=0.2\textwidth]{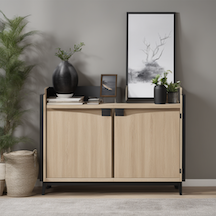}} & \raisebox{-\totalheight}{\includegraphics[width=0.2\textwidth]{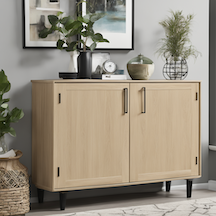}}  & \raisebox{-\totalheight}{\includegraphics[width=0.2\textwidth]{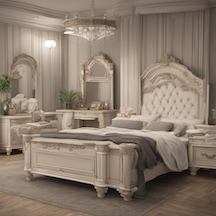}} \\
		MUZZ Sectional Sofa with Movable Ottoman, Free Combination Sectional Couch, Small L Shaped Sectional Sofa with Storage Ottoman, Modern Linen Fabric Sofa Set for Living Room (Dark Grey) & \raisebox{-\totalheight}{\includegraphics[width=0.2\textwidth]{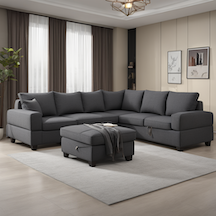}} & \raisebox{-\totalheight}{\includegraphics[width=0.2\textwidth]{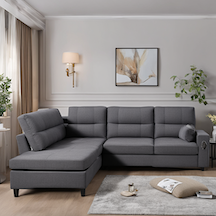}}  & \raisebox{-\totalheight}{\includegraphics[width=0.2\textwidth]{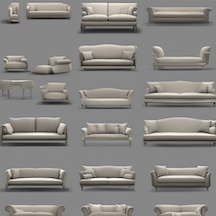}} \\
		Sony PS5 DualSense Wireless Controller - Midnight Black & \raisebox{-\totalheight}{\includegraphics[width=0.2\textwidth]{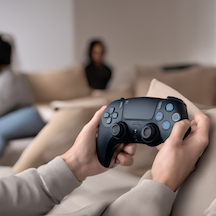}} & \raisebox{-\totalheight}{\includegraphics[width=0.2\textwidth]{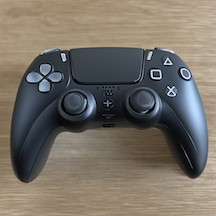}}  & \raisebox{-\totalheight}{\includegraphics[width=0.2\textwidth]{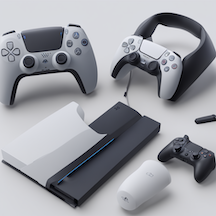}} \\
		Crayola Classic Crayons, Assorted Colors, Back to School, 24 Count & \raisebox{-\totalheight}{\includegraphics[width=0.2\textwidth]{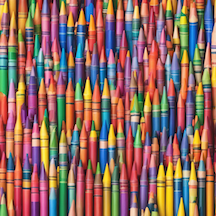}} & \raisebox{-\totalheight}{\includegraphics[width=0.2\textwidth]{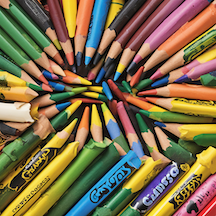}}  & \raisebox{-\totalheight}{\includegraphics[width=0.2\textwidth]{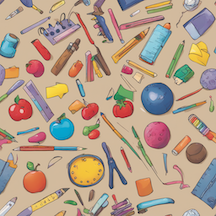}} \\
		\hline
	\end{tabular}
\end{table*}

To evaluate the relevance of the generated images concerning the product that a user has interacted with, we set up a human evaluation to rank the preference order of the generated images. The evaluators are data scientists working in the personalization domain and quality analysts responsible for visual quality assessment (VQA). Each of the evaluators was given a set of product names along with three images generated using the aforementioned approaches. The evaluators were given three options to rate each image for a given product: 1) low relevance, 2) medium relevance, and 3) high relevance. Each image could be assigned one of three ratings independent of the other images, that is, all three images of a given product could be rated as high(low) relevance at the same time. 

We collate the results from the human evaluation as follows. Twenty-four participants rated 15 product examples, ranging from clothing, pet, toy, and home decor categories. We map the relevance to the score as follows: low – 1, medium – 2, high – 3. Let $s_{i,j,k}$ be the score obtained for method $i$, product example $k$, and evaluator $j$. Let $I$ be the set of all the methods, that is, I = $\{\texttt{LLM}, \texttt{PNAME}, \texttt{PTYPE}\}$. Similarly, let $J$ be the set of all human evaluators and $K$ be the set of all product examples. The final score for a given method, say $i = \texttt{LLM}$, is given as follows: 

\begin{equation}\label{score_eq}
S_{\texttt{LLM}} = \frac{1}{|J| |K|} \sum_{j \in J} \sum_{k \in K} s_{i,j,k} \quad  i = \texttt{LLM}    
\end{equation}
where, $|\cdot|$ denotes the cardinality of a set.

\begin{table}
\centering 
\caption{The scores for human evaluation for three approaches: \texttt{LLM}, \texttt{PNAME}, and \texttt{PTYPE}.}\label{score_table} 
\begin{tabular} { c  c  c } \hline
\textbf{Method} & \textbf{Mean score} & \textbf{Std. dev.}\\ \hline 
\texttt{LLM} & 2.077 & 0.834\\ 
\texttt{PNAME} & 2.413 & 0.771 \\ 
\texttt{PTYPE} & 1.227 & 0.555\\ \hline
\end{tabular} 
\end{table}

The scores for each product are plotted in Figure \ref{fig:score_plot}. We observe the following from the scores: The \texttt{PTYPE} approach consistently performs worse than the other two approaches---\texttt{PNAME} and \texttt{LLM}. Between \texttt{PNAME} and \texttt{LLM}, we do not have a clear consensus among the participants on which one is generating more relevant images. For some products, the mean score for \texttt{PNAME} is higher whereas \texttt{LLM} has a larger mean score for others. If we average the scores across all products, as shown in Table \ref{score_table}, one can say that on average  \texttt{PNAME} is doing better than \texttt{LLM}; however, the large standard deviation does not warrant us to draw a clear conclusion. 

The generated images for the three approaches along with their product name for a sample of examples are available in Table \ref{tab:images_text}; additional examples are available in Appendix \ref{additional_examples}. We present an in-depth analysis of the generated images along with the nuanced differences between the two best-performing approaches:

\begin{itemize}
    \item For product name \textit{Vibrant Life Luxe Cuddler Mattress Edition Dog Bed, Medium, 27x21, Up to 40lbs} we obtain high-quality images of dog beds with cute dogs in a living room setting for both. However, the one produced using the chaining approach had the dog bed occupying a larger portion of the picture. This is because the LLM returned the prompt \textit{Cozy luxe cuddler mattress dog bed for medium-sized pups, up to 40lbs, available in 27"x21" size} emphasizing the token 27"x21" as a size feature which the product name itself didn't. In the case of \textit{Walker Edison 32" Scandinavian 2 Door Accent Cabinet - Coastal Oak/ Black}, the generated prompt \textit{Scandinavian-style 2-door accent cabinet in coastal oak and black, perfect for adding a touch of elegance to your home office or living room.} overlooked the feature of size and the product did not dominate the primary visual space of the image. Consequently, it ended up with an overall lower relevance score. 
    \item For product name \textit{MUZZ Sectional Sofa with Movable Ottoman, Free Combination Sectional Couch, Small L Shaped Sectional Sofa with Storage Ottoman, Modern Linen Fabric Sofa Set for Living Room (Dark Grey)}, the LLM returned the prompt \textit{Modern linen fabric sofa set with movable ottoman and storage, ideal for small living rooms in dark grey}. The image derived from \texttt{PNAME} based model fails to incorporate an Ottoman despite its repeated references. This could potentially be attributed to the model's inability to effectively process and distinguish among the multitude of keywords present.

    \item For product name \textit{Crayola Classic Crayons, Assorted Colors, Back to School, 24 Count}, the LLM created the prompt \textit{Colorful and versatile crayons perfect for back to school, ideal for children's creative expression in the classroom or at home}. The \texttt{PNAME} model focused on the brand name and item count which created an image with illegible text whereas the elimination of non-essential keywords resulted in a clearer image in the other one.
    
    \item For product name \textit{Sony PS5 Dualsense Wireless Controller - Midnight Black}, the LLM provided \textit{Wireless Sony PS5 Dualsense Controller in Midnight Black for Gaming on Your Couch}. The \texttt{LLM} based image receiving a lower relevance based on our survey results proved to be quite intriguing, as our initial marketing perspective led us to hypothesize that images portraying a person engaging with the controller, would be perceived as more relevant. However, the survey results contradicted our preconceived notion. This unexpected divergence from our initial assumption underscores the necessity for a more comprehensive evaluation. 

    \item In numerous examples, we noted that the product names consisted of multiple subjects resulting in differing outputs. For example, \textit{Dasein Women Satchel Handbags Top Handle Purse Medium Tote Bag Vegan Leather Shoulder Bag} was tagged by the LLM to be about a Tote Bag while the SD model considered it to be a Handbag resulting in different images. (see Appendix \ref{additional_examples} for details)
\end{itemize}

Given the above observations, we clearly see an opportunity for improvement. For instance, the product name is not always coherent, and relying solely on it to extract meaningful information is error-prone. We need to augment the meta-information of the product to exactly decipher the subject for creating the image. Secondly, the underlying LLM model does not always adhere to the given prompt, resulting in loss of information. This problem can be resolved using a better-performing LLM. Furthermore, a larger scale assessment would be needed to further explore and understand the relevance of such images in relation to online traffic patterns. 

\section{Conclusion and Future Work}
Generative AI has opened up many new avenues for businesses to improve user experience. One such avenue is creating personalized web banners. The current approach for generating web banners is time-consuming and does not scale well for a large number of users or user cohorts. Hence, there is a need to develop an automated banner generation technique that can be implemented fast and takes user preference into account.

In this work, we propose a technique to chain a large language model with a text-to-image model to automate the process of banner generation. The proposed approach uses LLM to extract important attributes from product names. The attributes are then passed to the image generation model to generate the banner image. We evaluate the quality of the generated images using the BRISQUE metric and find the results to be of high quality. In addition, we set up a user study to gauge the relevance of the generated banner with the product name whose attributes were used to generate the banner. The results of the user study indicate that the proposed approach is consistently generating images of medium to high relevance with the underlying product that guided the generation.

This work paves a path toward utilizing the power of generative AI models for automated and personalized web banner generation. Since this work is a pilot study, we limited the scope of evaluation to a limited number of products and used only the product name for attribute extraction. An immediate future direction for this work would be to include additional meta-information for a product while generating the attributes for capturing more granular differences. Furthermore, another worthy research direction is to utilize multiple products to extract a coherent prompt focused on always capturing specific attributes for banner generation. 

\begin{acks}
To all the participants of the user study who took time out from their busy schedules and helped us evaluate the generated images.
\end{acks}

\bibliographystyle{ACM-Reference-Format}
\bibliography{image_gen}

\newpage
\appendix

\section{Prompt to LLM}
\label{llm-prompt}
\begin{verbatim}
<<SYS>> 

As an English speaking ecommerce catalog \
specialist, you are given the product name. \
From this product name, extract a tuple \
containing the following elements: \

1. What the actual product is (Subject) 

2. What are the specific features of the \
product including the Brand, Color, or any \
Attributes (Keywords) 
3. What is the setting in which the product \
could be used. 

Include any entity that might use the \
product. (Setting) 

Convert the tuple to the following output - \
Subject with Keywords in Setting. \
Your response is limited to one sentence only. \
Remove all emojis and numbers from output. \
Only English alphabet characters are allowed. 

Sample Input: 

Nefoso Shag Light Gray Area Rug, 8' x 10' \
Soft Fluffy Area Rugs for Living Room Bedroom \
Kids Room Decor Carpet, Light Gray 


Sample output:  

fluffy and light gray area rug decorating \
the living room 

<</SYS>> 


[INST] 

User: A customer input the name \
'Peppa Pig 8-Inch Bean Plush Peppa Pig, \
Super Soft & Cuddly Small Plush Stuffed \
Animal, Kids Toys for Ages 2': 

What is the response for this product name? 

[/INST] 
\end{verbatim}
\onecolumn
\section{Additional examples from image generation survey}\label{additional_examples}
\begin{longtable}{p{0.195\textwidth} c c c}
\caption{Additional examples from image generation survey} \label{tab:additional_images} \\
\hline
Product Information & \texttt{LLM} & \texttt{PNAME} & \texttt{PTYPE} \\
\hline
Nefoso Shag Light Gray Area Rug, 8' x 10' Soft Fluffy Area Rugs for Living Room Bedroom Kids Room Decor Carpet, Light Gray &  \raisebox{-\totalheight}{\includegraphics[width=0.2\textwidth]{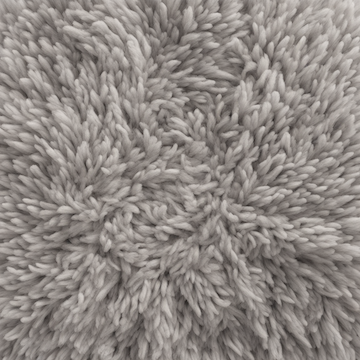}} & \raisebox{-\totalheight}{\includegraphics[width=0.2\textwidth]{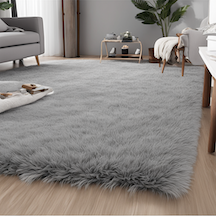}}  & \raisebox{-\totalheight}{\includegraphics[width=0.2\textwidth]{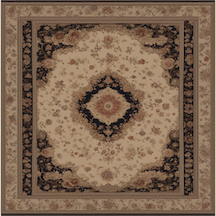}} \\
My Sweet Love 13 inch Baby Doll with Carrier and Handle Play Set, Light Skin Tone, Pink Theme &  \raisebox{-\totalheight}{\includegraphics[width=0.2\textwidth]{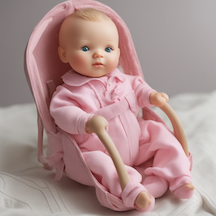}} & \raisebox{-\totalheight}{\includegraphics[width=0.2\textwidth]{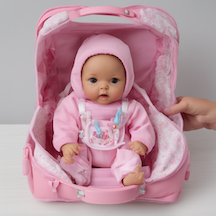}}  & \raisebox{-\totalheight}{\includegraphics[width=0.2\textwidth]{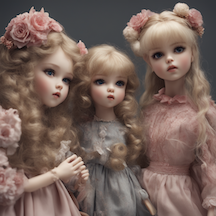}} \\
LEGO Heart Ornament Building Toy Kit, Heart Shaped Arrangement of Artificial Flowers, Great Gift for Valentine's Day, Unique Arts \& Crafts Activity for Kids, Girls and Boys  &  \raisebox{-\totalheight}{\includegraphics[width=0.2\textwidth]{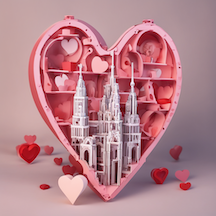}} & \raisebox{-\totalheight}{\includegraphics[width=0.2\textwidth]{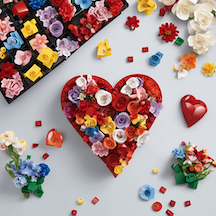}}  & \raisebox{-\totalheight}{\includegraphics[width=0.2\textwidth]{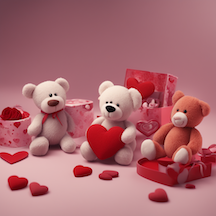}} \\
Peppa Pig 8-Inch Bean Plush Peppa Pig, Super Soft \& Cuddly Small Plush Stuffed Animal, Kids Toys for Ages 2 up &  \raisebox{-\totalheight}{\includegraphics[width=0.2\textwidth]{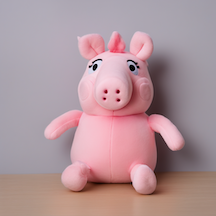}} & \raisebox{-\totalheight}{\includegraphics[width=0.2\textwidth]{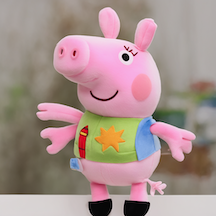}}  & \raisebox{-\totalheight}{\includegraphics[width=0.2\textwidth]{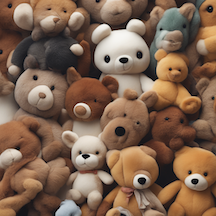}} \\
JETSTREAM 20" Hardside Spinner Rolling Carry-on Luggage, Durable ABS, with 2pcs Packing Cubes, Green &  \raisebox{-\totalheight}{\includegraphics[width=0.2\textwidth]{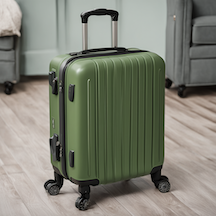}} & \raisebox{-\totalheight}{\includegraphics[width=0.2\textwidth]{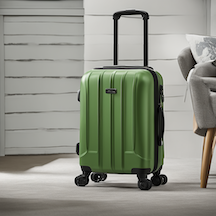}}  & \raisebox{-\totalheight}{\includegraphics[width=0.2\textwidth]{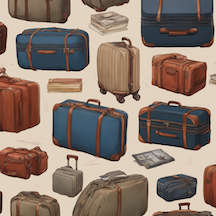}} \\
MTV Global Domination Men's and Big Men's Long Sleeve Graphic T-shirt &  \raisebox{-\totalheight}{\includegraphics[width=0.2\textwidth]{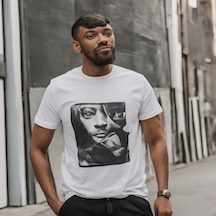}} & \raisebox{-\totalheight}{\includegraphics[width=0.2\textwidth]{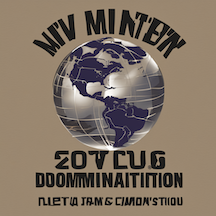}}  & \raisebox{-\totalheight}{\includegraphics[width=0.2\textwidth]{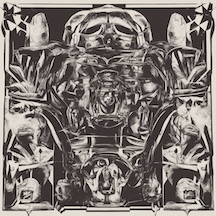}} \\
Kid's Basketball Shoes Boys Sneakers Girls Trainers Comfort High Top Basketball Shoes for Boys(Little Kid/Big Kid) White Black &  \raisebox{-\totalheight}{\includegraphics[width=0.2\textwidth]{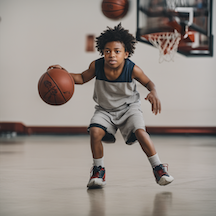}} & \raisebox{-\totalheight}{\includegraphics[width=0.2\textwidth]{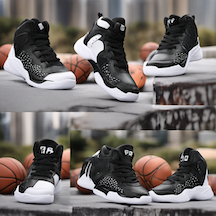}}  & \raisebox{-\totalheight}{\includegraphics[width=0.2\textwidth]{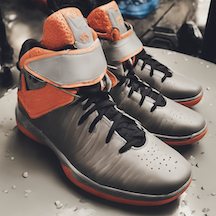}} \\
Lacoste Brown Shaded Rectangular Unisex Sunglasses L162S 210 61 &  \raisebox{-\totalheight}{\includegraphics[width=0.2\textwidth]{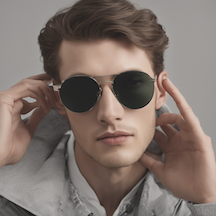}} & \raisebox{-\totalheight}{\includegraphics[width=0.2\textwidth]{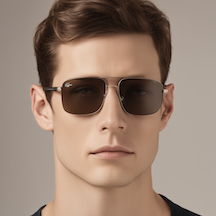}}  & \raisebox{-\totalheight}{\includegraphics[width=0.2\textwidth]{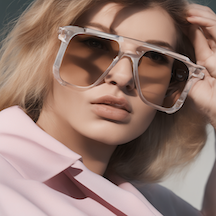}} \\
Dasein Women Satchel Handbags Top Handle Purse Medium Tote Bag Vegan Leather Shoulder Bag &  \raisebox{-\totalheight}{\includegraphics[width=0.2\textwidth]{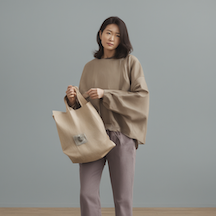}} & \raisebox{-\totalheight}{\includegraphics[width=0.2\textwidth]{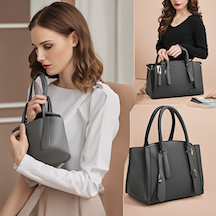}}  & \raisebox{-\totalheight}{\includegraphics[width=0.2\textwidth]{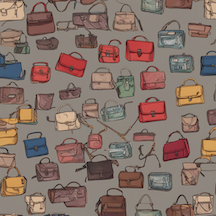}} \\
George Men's Baseball Hat &  \raisebox{-\totalheight}{\includegraphics[width=0.2\textwidth]{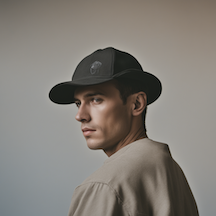}} & \raisebox{-\totalheight}{\includegraphics[width=0.2\textwidth]{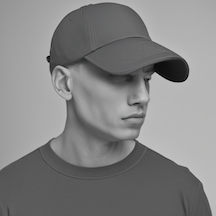}}  & \raisebox{-\totalheight}{\includegraphics[width=0.2\textwidth]{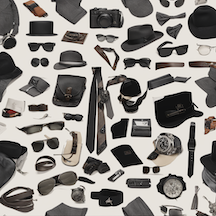}} \\
\hline
\end{longtable}

\end{document}